\documentstyle[11pt]{article}
\begin{document}
\centerline{\bf  DYNAMICAL GROUP FOR THE QUANTUM HALL 
EFFECT}
\centerline{}
\centerline{}
\centerline{Allan I. Solomon\footnote[1]{\noindent Permanent 
Address: Faculty of Mathematics and Computing, The Open 
University, Milton Keynes MK7 6AA,UK. E-mail:a.i.solomon@open.ac.uk}
}
\centerline{Laboratoire de Gravitation et Cosmologie Relativistes}
\centerline{Universit\'e Pierre et Marie Curie-CNRS}
\centerline{URA 769, Tour 22, 4\`eme \'etage, boite 142,}
\centerline{4, place Jussieu, 75252 Paris Cedex 05, France}
\centerline{ and }
\centerline{
 Joseph L. 
Birman\footnote[2]{\noindent Permanent Address: Department of 
Physics, City College of the City University of New York, New York 
10031.  Email:birman@sci.cuny.edu}}
\centerline{Department of Theoretical Physics}
\centerline{University of Oxford, UK}
\date{\today}
\begin{abstract}
A dynamical group for the single-particle (non-interacting)  Quantum Hall Effect is found, and is used to describe the Landau levels and determine the transverse (Hall) current.
\end{abstract}
\section{Introduction}
An area of  intense interest recently, both experimentally and 
theoretically, has been the Quantum Hall Effect. (See, for example, 
Reference \cite{qhe} and \cite{stone} for a review and reprint 
compilation on the Integral and Fractional effects.)  A rigorous 
treatment would require a microscopic interacting many-electron 
approach. Nevertheless, many features of the  effect may be 
observed in a classical treatment (the Hall current), or a 
single-particle quantum treatment (the Landau levels).  In this note we 
describe a dynamical group $G$ for the latter case.  Features 
of this group $G$ will remain in the interacting case, for which 
it must be the appropriate dynamical group in the noninteracting limit.  
In the absence of an electric field, the Landau levels appear 
naturally as the coherent states associated with $G$, while in the 
presence of an electric field $E$, the transverse Hall current arises 
naturally as an expectation in $E$-coherent states. 
\section{The dynamical group:  $E=0$ case}
The hamiltonian $H_M$ for an electron in a 
magnetic field is
\begin{equation}
H_M = \frac{1}{2m} ({\bf p}+ e{\bf A})^2\; \; \; \; (c=1)
\label{ham0}
\end{equation}
which, for a constant magnetic field $B$ along the $z$-axis,   may be written as 
\begin{equation}
H_M = \frac{1}{2m} (p_x - \frac{eB}{2} y)^2+\frac{1}{2m} (p_y + 
\frac{eB}{2} x)^2
\label{ham}
\end{equation}
or equivalently
\[
H_M = \frac{1}{2m} ({p_x}^2  + {p_y }^{2}) +\frac{1}{2} K (x^2 + y^2) 
+\frac{1}{2} \Omega L_z \]
\[ (\Omega = eB /m,\; K = e^2 B^2 /4m,\; \hbar=1) \]
where we have chosen the symmetric gauge 
\begin{equation}
{\bf A} = (-\frac{1}{2} B y, \frac{1}{2} B x,0)
\end{equation}
and the angular momentum operator $L_z$ about the $z$-axis is 
given by
\begin{equation}
 L_z = (x p_y -y p_x).
\end{equation}
Introduce boson operators $a_1, a_2$ in the usual way \cite{kahn}
\begin{eqnarray}
a_1 &=& (\frac{1}{2\ell})x + i \ell p_x \\
a_2 &=& (\frac{1}{2\ell}) y + i \ell p_y
\end{eqnarray}
\[ (\ell \equiv (eB)^{-\frac{1}{2}})\]
which satisfy
\begin{equation}
[a_1, {a_1}^{\dagger}] = 1, \; \; \; [a_2, {a_2}^{\dagger}] = 1, \; \; \; 
[a_1,a_2]=0.
\end{equation}
In terms of these operators, the hamiltonian Eqn.(\ref{ham}) 
becomes
\begin{eqnarray}
H_M &=& \frac{1}{2} \Omega ({a_1}^{\dagger} a_1 + \frac{1}{2}) 
+\frac{1}{2} \Omega ({a_2}^{\dagger} a_2+ \frac{1}{2}) + 
\frac{1}{2} \Omega i(a_1 a_2^{\dagger}-a_2 a_1^{\dagger}) 
\nonumber \\
&=& \frac{1}{2} \Omega ({a_1}^{\dagger} a_1 + {a_2}^{\dagger} 
a_2+ 1) + \frac{1}{2} \Omega L_z
\label{ham2}
\end{eqnarray}
where the angular momentum operator $L_z$ is 
\begin{equation}
L_z = i(a_1 a_2^{\dagger}-a_2 a_1^{\dagger}) .
\label{eq:ang}
\end{equation}
The hamiltonian Eqn.(\ref{ham2}) is an element of the Lie algebra 
$u(2)$ generated by 
$\{ a_i^{\dagger} a_j  : (i,j)=1,2 \}$, which is therefore its 
spectrum-generating algebra\cite{yuval} (SGA) (or dynamical 
algebra).  This is the case where there is  no electric field $E$ 
present. Additionally, we emphasize that in the present context 
we are taking a single-electron viewpoint;  for a system of {\em 
non-interacting} electrons, the SGA would be a direct sum 
$\bigoplus u(2)$. Note that the {\em symmetry} group is the 
$SO(2)$ of rotations about the $z$-axis generated by $L_z$, since 
$[L_z,H_M]=0.$ 

Even at this elementary level, the SGA enables a simple  and 
immediate interpretation of the Landau levels associated with the 
hamiltonian Eqn.(\ref{ham}).  
Choose a conventional $u(2)$ basis  
\begin{equation}
\{ J_0, J_1, J_2, J_3 \} = \{ \frac{1}{2}(n_1+n_2), \; 
\frac{1}{2} (a_1 a_2^{\dagger}+a_2 a_1^{\dagger}) , \; 
  \frac{i }{2} (a_1 a_2^{\dagger}-a_2 a_1^{\dagger}), \; 
  \frac{1}{2} (n_1-n_2) \}
\end{equation}
and for brevity define \begin{equation}
A^{\dagger} \equiv (a{_1}^{\dagger},a{_2}^{\dagger}).
\label{eq:A}
\end{equation}
We may represent a typical (Bogoliubov) rotation by
\begin{equation}
R^{\dagger}_{k}(\theta)  A R_{k}(\theta) = \exp (i \theta 
\sigma_k) A
\end{equation}
where 
\begin{equation}
\{\sigma_1, \; \sigma_2, \;  \sigma_3 \} =  \{ \frac{1}{2} \left[ 
\begin{array}{cc} 0 & 1 \\  1 & 0 \end{array} \right],  \; \; 
\frac{1}{2}\left[ \begin{array}{cc} 0 & -i \\ i & 0 \end{array} 
\right], \; \; \frac{1}{2}\left[ \begin{array}{cc} 1 & 0 \\ 0 & -1 
\end{array} \right] \}
\end{equation}
\begin{equation}
\{ R_{1}(\theta), \; R_{2}(\theta), \; R_{3}(\theta) \} =  \{ e^{(i J_1 
\theta)} , \; \; e^{(i J_2 \theta) }, \; \; 
e^{(i J_3 \theta)} \}
\end{equation}
The energy eigenvalue equation for the hamiltonian $H_M$ as
\begin{equation}
H_M \psi(n_1,n_2) = E(n_1,n_2) \psi(n_1,n_2)
\end{equation}
and  $H_M$ is represented (apart from the $\frac{1}{2} \Omega$ 
constant additive term) by a  $2 \times 2$ matrix $M_M$ 
\begin{equation}
H_M = \frac{1}{2} \Omega A^{\dagger} M_M A
\end{equation}
with
\begin{equation}
M_M = \left[ \begin{array}{cc} 1 & -i \\ i & 1 \end{array} \right].
\end{equation}
In this basis the angular momentum operator $L_z$ 
Eqn.(\ref{eq:ang})
is represented by
\begin{equation}
M_L = \left[ \begin{array}{cc} 1 & 0 \\ 0 & -1 \end{array} \right].
\end{equation}

Diagonalisation of the hamiltonian corresponds to taking a unitary 
(Bogoliubov) transformation on $A$
\begin{equation}
A \rightarrow U^{\dagger} A U = D A
\end{equation}
(the matrix $D$ gives a realization of the operator $U$)
so that
\begin{equation}
H \rightarrow U^{\dagger} H U = H^{D}_{M}
\end{equation}
is diagonal (a function only of the number operators $n_1,n_2$), 
so that
\begin{equation}
H^{D}_M | n_1,n_2 \rangle = E(n_1,n_2) | n_1,n_2 \rangle 
\end{equation}
with 
\begin{equation}
\psi(n_1,n_2) = U | n_1,n_2 \rangle .
\end{equation}

Diagonalisation is effected by taking 
\begin{itemize}
\item Case (i) $U=R_1(\pi/2)$
\begin{eqnarray}
M_M & \rightarrow &  \left[ \begin{array}{cc} 2 & 0 \\ 0 
& 0 \end{array} \right]  \; \; H^{D}_M = \Omega (n_1+\frac{1}{2}) 
\\
M_L & \rightarrow &  \; \; \left[ \begin{array}{cc} 1 & 0 \\ 0 & -1 
\end{array} \right]  \; \; L^{D} = (n_1-n_2) 
\end{eqnarray}
\\
\item Case (ii) $U=R_1(-\pi/2) $
\begin{eqnarray}
M_M & \rightarrow & \left[ \begin{array}{cc} 0 & 0 \\ 0 
& 2 \end{array} \right]  \; \; H^{D}_M = \Omega (n_2+\frac{1}{2}) 
\\
M_L & \rightarrow &  \; \; \left[ \begin{array}{cc} -1 & 0 \\ 0 & 1 
\end{array} \right]  \; \; L^{D} = (n_2-n_1) \end{eqnarray}
\end{itemize}
In either case, there is a degeneracy, with only one eigenvalue 
labelling the energy levels (Landau levels).

The simultaneous eigenstates $\psi(n_1,n_2)$ of $H_M$ and $L_z$ 
are given by
\begin{eqnarray}
\psi(n_1,n_2) &=& R_1( \theta) |n_1, n_2 \rangle \nonumber \\
&=& \frac{1}{\sqrt{(n_1! n_2!)}}R_1( \theta) ({a_1}^{\dagger} 
)^{n_1}({a_2}^{\dagger})^{n_2}|0, 0\rangle \nonumber \\
&=& \frac{1}{\sqrt{(n_1! n_2!)}}R_1( \theta) ({a_1}^{\dagger} 
)^{n_1}({a_2}^{\dagger})^{n_2}R_1(\theta)^{\dagger}|0, 0\rangle 
\nonumber \\
&=& \frac{1}{\sqrt{(n_1! n_2!)}}(a_{1}^{\dagger}+i 
a_{2}^{\dagger})^{n_1}(a_{2}^{\dagger}+i a_{1}^{\dagger})^{n_2}|0, 
0\rangle
\label{eq:psi}
\end{eqnarray}
using
\begin{equation}
R_1(\theta) A R{_1}^{\dagger}(\theta) = e^{-i \theta {\sigma}_{1} } 
A\; \; \;  \; \;  R_1(\theta) = e^{i \theta J_1}
\end{equation}
so that 
\begin{equation}
a_1 \rightarrow (a_1-i a_2)/\sqrt{2} \; \; \; \; \; \; a_2 
\rightarrow (a_2-i a_1)/\sqrt{2} \; \; \; \; \; \;  ( \theta =\pi / 2)
\end{equation}
The eigenstates Eqn.(\ref{eq:psi})  are $u(2)$ coherent states. 
\section{The dynamical group:  $E\neq 0$ case}
We now introduce an electric field  ${\bf E} = (E_1,\;E_2,\;0)$ with 
associated term
\begin{eqnarray}
H_E &= & exE_1+eyE_2 \nonumber \\
&=&e\;E_1 \ell (a_1+a^{\dagger}_1)\; +\; e\;E_2 \ell 
(a_2+a^{\dagger}_2) 
\end{eqnarray}
The total hamiltonian $H$ in the presence of magnetic field ${\bf B}$ 
and  electric field ${\bf E}$ is given by 
\begin{eqnarray}
H &=& \frac{1}{2m} (p_x - \frac{eB}{2} y)^2+ \frac{1}{2m} (p_y + 
\frac{eB}{2} x)^2 +exE_1+eyE_2   \nonumber \\
 &=& \frac{1}{2} \Omega ({a_1}^{\dagger} a_1 + {a_2}^{\dagger} 
a_2+ 1) + \frac{i}{2} \Omega (a_1 a_2^{\dagger}-a_2 
a_1^{\dagger})+\frac{1}{2} \Omega ({\cal E}_1 
(a_1+{a_1}^{\dagger})  \nonumber  \\
& &+{\cal E}_2 (a_2+ {a_2}^{\dagger}))  
\label{eq:he}
\end{eqnarray}
in units such that $ {\cal E}_i \equiv \frac{2 e \ell}{\Omega} E_i$.
The SGA in this case is the 9-dimensional algebra ${\cal L}$ 
generated by 
$\{ a_i^{\dagger} a_j \; a_i^{\dagger} , a_j , I: (i,j)=1,2 \}$.
The structure of this algebra may be elucidated from a 
Levi-Malcev decomposition via its maximal solvable radical ${\cal N}$
\begin{equation}
{\cal N} = \{a_1,a_1^{\dagger}, a_2, a_2^{\dagger},n_1+n_2, I \} \; 
\; \; \; (n_i \equiv a_i^{\dagger}a_i)
\end{equation}
with
\begin{equation}
{\cal L}/{\cal N} = \{ a_1 a_2^{\dagger}, a_2^{\dagger} a_1, n_1-
n_2\} =su(2)
\end{equation}
giving ${\cal L} = su(2){\bigcirc  \! \! \! \!s} \, {\cal N}.$ The 
corresponding group ${\cal G}$ is generated by the unitary actions 
on ${\cal L} $ of the form $y \rightarrow e^x y e^{x^{\dagger}}$ 
where $x,y$ are (anti-hermitian) elements of ${\cal L} $.
These unitary actions are:
\begin{itemize}
\item Rotations of the form 
\begin{eqnarray}
a_1 &\rightarrow& \lambda a_1 + \mu a_2  \nonumber \\
a_2 &\rightarrow& -\overline{{\mu}} a_1 + 
\overline{{\lambda}}a_2  \; \; \; (|\lambda |^2+|\mu |^2=1),
 \label{eq:rot}
\end{eqnarray}
\item displacements 
\begin{eqnarray}
a_1 &\rightarrow& a_1 + {\lambda}_1  \nonumber \\
a_2&\rightarrow& a_2+ {\lambda}_2 
 \label{eq:dis}
\end{eqnarray}
\item and phase transformation
\begin{eqnarray}
a_1 &\rightarrow& e^{i \phi}a_1 \nonumber \\
a_2 &\rightarrow& e^{i \phi}a_2 . \label{eq:phase}
\end{eqnarray}
\end{itemize}
This gives rise to the 8-dimensional inhomogeneous unitary group 
$IU(2)$ (the Center of ${\cal L} $ does not contribute to the 
unitary actions).  By extending the definition of 
$A$  in Eqn.(\ref{eq:A})  to
\begin{equation}
A^{\dagger} \equiv (a{_1}^{\dagger},\; \; a{_2}^{\dagger},\; \; 1)
\label{eq:bigA}
\end{equation}
we may realize the dynamical  group (ignoring the phase transformation Eqn.(\ref{eq:phase}) above) by
\begin{equation}
G =\left\{ \left[ \begin{array}{ccc} 
\lambda & \mu & {\lambda}_1 \\ 
 -\overline{\mu} & \overline{\lambda} & {\lambda}_2 \\ 
0 & 0 & 1 \end{array} \right] {|} \; \;  |{\lambda}|^2+|\mu|^2 =1, \; \; \;  \lambda, \mu, {\lambda}_1,{\lambda}_2 \in {\cal C} \right\}.
\end{equation}
The hamiltonian Eqn.(\ref{eq:he}) may be written in this basis as 
$H=\frac{1}{2} \Omega A^{\dagger} M A$ where
\begin{equation}
 M =\left [ \begin{array}{ccc} 1 & -i & {\cal 
E}_1 \\  i & 1 & {\cal E}_2 \\  {\cal E}_1 &  {\cal E}_2 & 1 
\end{array} \right]
\end{equation}
for ${\cal E}_{i}$ real.
\section{The Hall current}
The hermitian hamiltonian Eqn.(\ref{eq:he})  may be diagonalised 
by a unitary transformation.  However, it is of value to obtain a 
canonical form under the transformations Eqn.(\ref{eq:rot}) and 
Eqn.(\ref{eq:dis}); this will enable us to obtain the coherent states 
of the system in the presence of an electric field, which play a role 
in the Hall effect. It is straightforward to prove that for singular 
$M_H$ (the case here) the matrix $M$ may not be sent to diagonal 
form under inner automorphisms of the algebra ${\cal L}$, that is, 
by transformations of the group $G$.  For example, choosing the 
electric field ${\bf E}$ along the $x$-axis ($E_2 = 0$), 
under the transformation,
\begin{equation}
{ M} \rightarrow T^{\dagger} {M} T
\label{trans}
\end{equation}
where $T$ represents the unitary transformation  $R_1(\frac{\pi}{2}) D({\lambda}_1,{\lambda}_2)$
for displacements 
\begin{equation}
{\lambda}_1 = -{{{\cal E}_{1}}\over 2{\sqrt{2}}} \; \; \; 
{\lambda}_{2} = -i{{{\cal E}_{1}}\over {4\sqrt{2}}} .
\end{equation}
the canonical form is
\begin{equation}
M \rightarrow\left[ \begin{array}{ccc}
2&0&0\\
  0&0&-i {{\cal E}_{1}}\over {\sqrt{2}} \\
0& i {{\cal E}_{1}}\over {\sqrt{2}}&1 
\end{array}\right].
\end{equation}
This corresponds to the Bogoliubov transformation in Fock space
\begin{eqnarray}
H &\rightarrow &U^{\dagger} H U \nonumber \\
    &=& \Omega ({a_1}^{\dagger} a_1 + \frac{1}{2}) + 
\frac{i \Omega}{2 \sqrt{2}} {\cal E}_{1} (a_2-{a_2}^{\dagger}).\\
\end{eqnarray}
Although this is not diagonal in the number operators, it is simply 
a sum of a commuting number operator and momentum
\begin{equation}
H \rightarrow  \Omega (n_1 +\frac{1}{2}) - \frac{\sqrt(2)E_x}{B} p_y
\end{equation}
from which the eigenstates $|n_1,k_y\rangle$ are immediate.

We now calculate the Hall current by this method.
The canonical current operator for an electron in the presence of the vector 
potential ${\bf A}$ is given by 
\begin{eqnarray} 
{\bf J}&=&\frac{e}{m}({\bf p}+e{\bf A})\nonumber \\
	&=&\frac{e}{m}((p_x - \frac{eB}{2} y), (p_y + \frac{eB}{2} 
x),0)\nonumber \\
	&=&\frac{e}{2m\ell}(i({a_1}^{\dagger} -{a_i})-(a_2+
{a_2}^{\dagger}),i({a_2}^{\dagger} -a_{2})-(a_i+{a_1}^{\dagger}), 0) \nonumber
\end{eqnarray}
which may be represented by
\begin{equation}
{\bf j} = j_0\left( \left[ \begin{array}{ccc} 0 & 0 & i \\ 0 & 0 & -1\\  -i &  -1 & 0 \end{array} \right] ,\left [ \begin{array}{ccc} 0 & 0 & 1
\\ 0 & 0 & i\\  1 &  -i & 0 \end{array} \right] ,0 \right) \; \; \; 
(j_0\equiv\frac{e}{2m\ell}).
\end{equation}
Under the transformation Eqn.~(\ref{trans}),
\begin{eqnarray}
{\bf j} &\rightarrow &T^{\dagger} {\bf j} T \nonumber \\
	&=& j_0\left( \left[ \begin{array}{ccc} 0 & 0 & \sqrt{2}i \\ 0 & 0 & 0\\  -\sqrt{2}i & 0 & 0 \end{array} \right] ,\left [ \begin{array}{ccc} 0 & 0 & \sqrt{2}
\\ 0 & 0 & 0\\  \sqrt{2} &  0 & -{\cal E}_{1} \end{array} \right] ,0 \right) 
\end{eqnarray}
This corresponds in Fock space to
\begin{equation}
{\bf J}\rightarrow \frac{e}{2m\ell}(\sqrt{2}i({a_1}^{\dagger} -{a_i}),
\sqrt{2}({a_1}^{\dagger} +{a_i})-{\cal E}_1, 0) 
\label{curr}
\end{equation}
and  shows immediately that the ground state current density (Hall current) is given by $-\frac{e}{2ml} {\cal E}_1\equiv {{-eE_x }\over {B}}$ in the $y$-direction. 
As is well-known, this  treatment gives no longitudinal current (drift current) \cite{stone},\cite{stone2} which arises, as is clear from Eqn.(\ref{curr}), from mixing of the Landau levels. 
\section{Acknowledgements}
The authors acknowledge the support of   NATO Grant Number  CRG 940452, 
under whose auspices this research was conducted. We also thank Professor Richard Kerner of  the University of  Paris, and Professors David Sherrington and Sir Roger Elliott of the University of Oxford, for hospitality. We thank Leonard Tevlin for pointing out reference\cite{stone2}  . \\ \\

\end{document}